\documentclass[lettersize,journal]{IEEEtran}
\usepackage{cite}
\usepackage{amsmath,amssymb,amsfonts}
\usepackage{graphicx}
\usepackage{textcomp}
\usepackage{algorithm,algorithmic}
\usepackage{multicol}
\usepackage{multirow}
\usepackage{changepage,threeparttable}
\usepackage{makecell}
\usepackage{array}
    \newcolumntype{P}[1]{>{\centering\arraybackslash}p{#1}}
    \newcolumntype{M}[1]{>{\centering\arraybackslash}m{#1}}
\usepackage{booktabs, multirow}
\usepackage{soul}

\usepackage{amsmath,amsfonts}
\usepackage{algorithmic}
\usepackage{array}
\usepackage[caption=false,font=normalsize,labelfont=sf,textfont=sf]{subfig}
\usepackage{textcomp}
\usepackage{stfloats}
\usepackage{url}
\usepackage{verbatim}
\usepackage{graphicx}
\def\BibTeX{{\rm B\kern-.05em{\sc i\kern-.025em b}\kern-.08em
    T\kern-.1667em\lower.7ex\hbox{E}\kern-.125emX}}
\usepackage{balance}

\begin{document}
\title{Fully Automated Assessment of Cardiac Coverage in Cine Cardiovascular Magnetic Resonance Images using an Explainable Deep Visual Salient Region Detection Model}
\author{Shahabedin Nabavi, Mohammad Hashemi, Mohsen Ebrahimi Moghaddam, Ahmad Ali Abin, \\Alejandro F. Frangi, \IEEEmembership{Fellow, IEEE}
\thanks{ }
\thanks{S. Nabavi, M. Hashemi, M. Ebrahimi Moghaddam and A. A. Abin are with the Faculty of Computer Science and Engineering, Shahid Beheshti University, Tehran, Iran (Corresponding Author: M. Ebrahimi Moghaddam, e-mail: m\underline{ }moghadam@sbu.ac.ir)
}
\thanks{A. F. Frangi is with the Centre for Computational Imaging and Simulation Technologies in Biomedicine (CISTIB), School of Computing, University of Leeds, UK. }
\thanks{}
}

\maketitle

\begin{abstract}
Cardiovascular magnetic resonance (CMR) imaging has become a modality with superior power for the diagnosis and prognosis of cardiovascular diseases. One of the essential basic quality controls of CMR images is to investigate the complete cardiac coverage, which is necessary for the volumetric and functional assessment. This study examines the full cardiac coverage using a 3D convolutional model and then reduces the number of false predictions using an innovative salient region detection model. Salient regions are extracted from the short-axis cine CMR stacks using a three-step proposed algorithm. Combining the 3D CNN baseline model with the proposed salient region detection model provides a cascade detector that can reduce the number of false negatives of the baseline model. The results obtained on the images of over 6,200 participants of the UK Biobank population cohort study show the superiority of the proposed model over the previous state-of-the-art studies. The dataset is the largest regarding the number of participants to control the cardiac coverage. The accuracy of the baseline model in identifying the presence/absence of basal/apical slices is 96.25\% and 94.51\%, respectively, which increases to 96.88\% and 95.72\% after improving using the proposed salient region detection model. Using the salient region detection model by forcing the baseline model to focus on the most informative areas of the images can help the model correct misclassified samples' predictions. The proposed fully automated model's performance indicates that this model can be used in image quality control in population cohort datasets and also real-time post-imaging quality assessments.
\end{abstract}

\begin{IEEEkeywords}
Cardiac Coverage, Cardiovascular magnetic resonance imaging, Image quality assessment, Salient region detection, Population image analysis.
\end{IEEEkeywords}

\section{Introduction}
\label{sec:introduction}
\IEEEPARstart{P}{opulation} imaging is performed to help preventive approaches implementation based on big data by collecting imaging data from large populations. Population imaging has become a major medical issue in recent years, shifting curative to preventive medicine. Preventive medicine can improve people's quality of life and reduce treatment costs \cite{attar2019quantitative}. Designing and implementing predictive models requires using robust data so that the results of these models can be trusted. Therefore, in the first step, the quality of imaging data should be assessed, which is a challenge in population imaging studies due to its expansion. Designing automated methods to evaluate image quality can ensure accuracy in predictive models and eliminate the time-consuming and costly subjective image quality assessment process \cite{chow2016review}.

UK Biobank is the most extensive population imaging study by collecting image data with different modalities of 100,000 participants \cite{littlejohns2020uk}. Despite the precise protocols used to collect the data \cite{petersen2015uk}, a part of the collected imaging data is accompanied by distortions that can adversely affect the results of studies defined based on this data. To identify suboptimal images and exclude them from further studies, it is necessary to propose automatic methods appropriate to the large volume of such cohorts. Cardiovascular magnetic resonance images (CMRIs) of the UK Biobank have been used to suggest some distinctive approaches to assessing the function and structure of the cardiovascular system. One of the significant problems that challenge the accuracy of these approaches to cardiac volumetric measurements is the absence of basal and/or apical slices in the short-axis CMR imaging \cite{klinke2013quality}. Therefore, we propose a method to automatically examine cine CMR images for complete cardiac coverage and identify stacks with missing basal/apical slices. For this purpose, a deep 3D convolutional neural network (CNN) is proposed, and the accuracy is then improved by using a novel explainable salient region detection model.

The contributions of this study are:
\begin{enumerate}
    \item An automated approach for image quality assessment of short-axis cine CMR images is proposed to detect the absence of basal/apical slices in these images. This approach uses a deep 3D CNN architecture, which is improved by determining the salient regions of the images. It is shown that by modifying the input images based on the detection of salient regions, the neural network training process can be guided to improve accuracy.
    \item A novel learning-based model is proposed for image salient region detection. The proposed interpretable model can detect the 3D image salient regions so that this interpretability increases the reliability of the model. In this way, a local interpretable model-agnostic explanations technique interprets the deep CNN task. Then, the salient regions are extracted by an attention-based 3D U-Net network to obtain the final explainable salient region detection model.
\end{enumerate}

The remainder of this article is organised as follows. Section II reviews related studies. Section III describes preparing the training and testing sets and the proposed models. Section IV provides experimental setups and metrics. The results of the experiments are presented in Section V. Section VI discusses the results obtained, followed by the conclusion in Section VII.

\section{Related Works}
To increase the accuracy of functional and volumetric measurements of the heart, it is necessary to have complete coverage of the left ventricle (LV) from the basal to the apex in short-axis cine CMR images \cite{petersen2017reference}. Although there are specific guidelines for determining reliable margins in CMR imaging \cite{schulz2020standardized}, some imaging volumes may lack sufficient information for image analysis. The most significant limitation for complete LV coverage is the absence of basal and/or apical slices. The missing basal slice makes the atrial chamber at the end-systole invisible, so the base of the heart is not entirely covered in the image. The absence of an apical slice also causes the LV cavity to remain visible at the end-systole, making it challenging to determine LV volume \cite{klinke2013quality}. Studies in recent years have provided automated methods for examining complete left ventricular coverage, which we will review in the following.

In two studies by Tarroni et al. \cite{tarroni2020large, tarroni2018learning}, complete heart coverage was investigated using a hybrid decision forest method. The evaluation results of this method on the images related to nearly 3,000 cases from the UK Biobank and 100 cases from the UK Digital Heart Project show that the proposed method has 88\% sensitivity and 99\% specificity. In a study by Zhang et al. \cite{zhang2017semi}, A method called Semi-Coupled-GANs (SCGANs) was used to investigate the presence or absence of basal and apical slices in short-axis cine CMR volumes. About 90\% accuracy has been reported on a database containing over 6,000 samples from the UK Biobank for this study. In \cite{zhang2018multi}, A dataset-invariant adversarial learning framework for detecting basal and apical slices is presented. The proposed approach is based on dataset adaption, and three datasets have been used for evaluation, with the highest accuracy of about 92\%. In another study by Zhang et al. \cite{zhang2018automatic}, A modified version of 3D CNN called fisher-discriminative 3D CNN has been used to detect missing basal/apical slices. In this study, an attempt has been made to minimize within-class scatter and maximize between-class scatter by changing the loss function. The results on over 5,000 volumetric scans of short-axis cine CMR images from the UK Biobank show a precision of about 91\%.

Although, according to the literature \cite{klinke2013quality}, investigation of complete LV coverage is the first criterion for CMR image quality assessment, other studies have examined other artefacts. Some studies have analyzed the motion artefacts in these images \cite{tarroni2020large,tarroni2018learning,oksuz2019automatic}. Our previous study \cite{nabavi2021automatic} proposed an automated method for identifying four common artefacts in CMR images in spatial and frequency domains. A domain adaptation approach was used to make the proposed method more generalizable, and the proposed method was evaluated on four different datasets, including the UK Biobank.

\section{Method}
In this section, we first discuss how to prepare data for training and application of the model and then describe the proposed method in detail. The proposed deep convolutional learning model for detecting the presence or absence of basal/apical slices is described. Besides, an attempt is made to improve the accuracy of the deep convolutional model by proposing a new interpretable salient region detection method.

\subsection{Preparing Training and Testing Image Stacks}
\begin{figure}[!t]
\centerline{\includegraphics[width=\columnwidth]{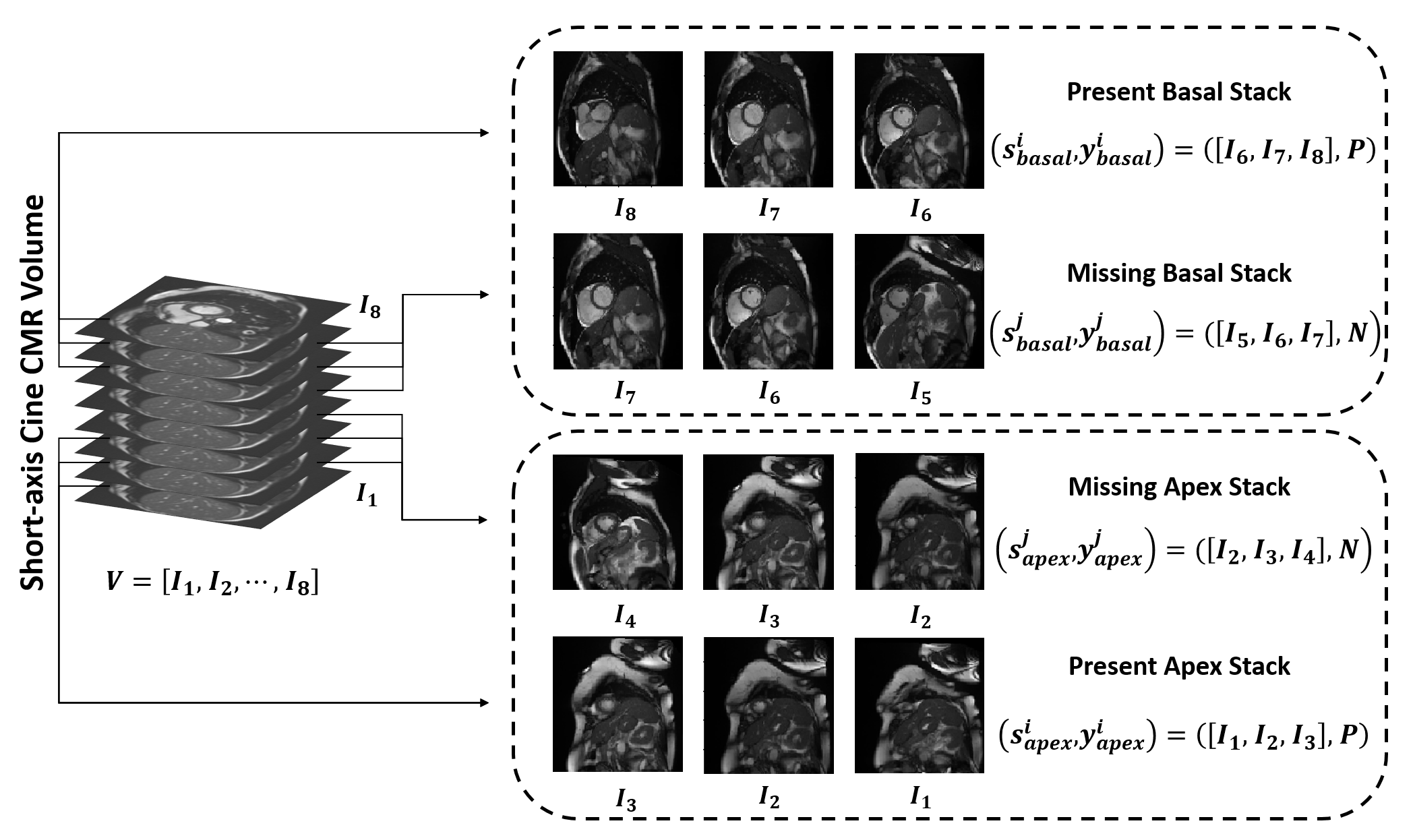}}
\caption{How to prepare the positive/negative triplets to train and test the proposed model for detecting the presence/absence of basal/apical slices.}
\label{fig1}
\end{figure}

By examining the short-axis cine CMR images of over 6,200 cases from the UK Biobank, imaging stacks of 5,449 cases with full cardiac coverage were selected for training and testing. Full-coverage stacks typically include 8 to 10 slices, covering from basal to apex. The top three slices are extracted from the main stacks and used as a positive triplet to detect the presence of basal. The bottom three slices of the stack, the last of which is the apex slice, are selected and stored as a positive triplet for the apex presence. By excluding basal and apical slices from the main stacks, the successive three slices from the top and the next three slices from the bottom are considered negative triplets related to the absence of basal and apex, respectively. With the mentioned method for preparing training and testing dataset, a total of 10,898 stacks, including positive and negative samples, were considered to evaluate the presence or absence of basal, and the same number to evaluate the presence or absence of apex. Fig. \ref{fig1} shows how to prepare the datasets mentioned above.

\subsection{The Proposed Cardiac Coverage Assessment Method}
In this section, we describe the proposed model for cardiac coverage assessment. At first, the problem formulation is presented, and then the baseline model to assess the complete coverage of the heart is described. After that, we examine the interpretability of the baseline model, and with a U-Net-based learning model, the salient region of the input cine CMR stacks are extracted. Finally, a cascaded setting is proposed that can improve the performance of the baseline based on the extracted salient regions.

\subsubsection{Problem Formulation}
We use vector \(V\) to express a full-coverage stack. Each stack consists of several images, usually ranging from \(8\) to \(10\).

\begin{equation}
V=[I_1, I_2,\cdots, I_n], n\in \left\{{8}, {9}, {10}\right\}
\label{eq1}
\end{equation}

The training and testing stacks must now be prepared as described in the previous section. Two datasets are created, one for the positively and negatively labelled apical stacks \((S_{apex})\) and one for the positively and negatively labelled basal stacks \((S_{basal})\) for the two tasks of identifying the presence/absence of basal/apical slices, respectively (Refer to \eqref{eq2} and \eqref{eq3}). Let \((s^i, y^i)\) be a training sample, where \(s^i\) is the \(i^{th}\) stack and \(y^i\) represents the label for \(s^i\). Labels include positive \((P)\) and negative \((N)\) values.

\begin{equation}
S_{apex}=\{(s^1_{apex}, y^1_{apex}), \cdots, (s^n_{apex}, y^n_{apex})\}, y \in \{P, N\}
\label{eq2}
\end{equation}

\begin{equation}
S_{basal}=\{(s^1_{basal}, y^1_{basal}), \cdots, (s^n_{basal}, y^n_{basal})\}, y \in \{P, N\}
\label{eq3}
\end{equation}

According to \eqref{eq2} and \eqref{eq3}, stacks formed based on vector \(V\) can be divided into four categories mentioned in \eqref{eq4} and \eqref{eq5}.

\begin{equation}
\left\{
	\begin{array}{lll}
		(s^i_{apex}, y^i_{apex})=([I_1, I_2, I_3], P)  &  \\
		 & i \neq j\\
		(s^j_{apex}, y^j_{apex})=([I_2, I_3, I_4], N)  &
	\end{array}
\right.
\label{eq4}
\end{equation}

\begin{equation}
\left\{
	\begin{array}{lll}
		(s^i_{basal}, y^i_{basal})=([I_{n-2}, I_{n-1}, I_{n}], P)  &  \\
		 & i \neq j,\\
		 & n\in \left\{{8}, {9}, {10}\right\}\\
		(s^j_{basal}, y^j_{basal})=([I_{n-3}, I_{n-2}, I_{n-1}], N)  &
	\end{array}
\right.
\label{eq5}
\end{equation}

These stacks are then used to train two deep convolutional neural network models. In the first model, the training is done on dataset \(S_{apex}\) to detect the presence/absence of apical slices. In the second model, the training is performed on dataset \(S_{basal}\) to identify the presence/absence of basal slices. Finally, the trained networks' weights related to datasets \(S_{apex}\) and \(S_{basal}\) are stored as \(\omega_{apex}\) and \(\omega_{basal}\), respectively.

\subsubsection{3D Convolutional Baseline Model for Cardiac Coverage Assessment}

\begin{figure*}[!t]
\centerline{\includegraphics[width=\textwidth]{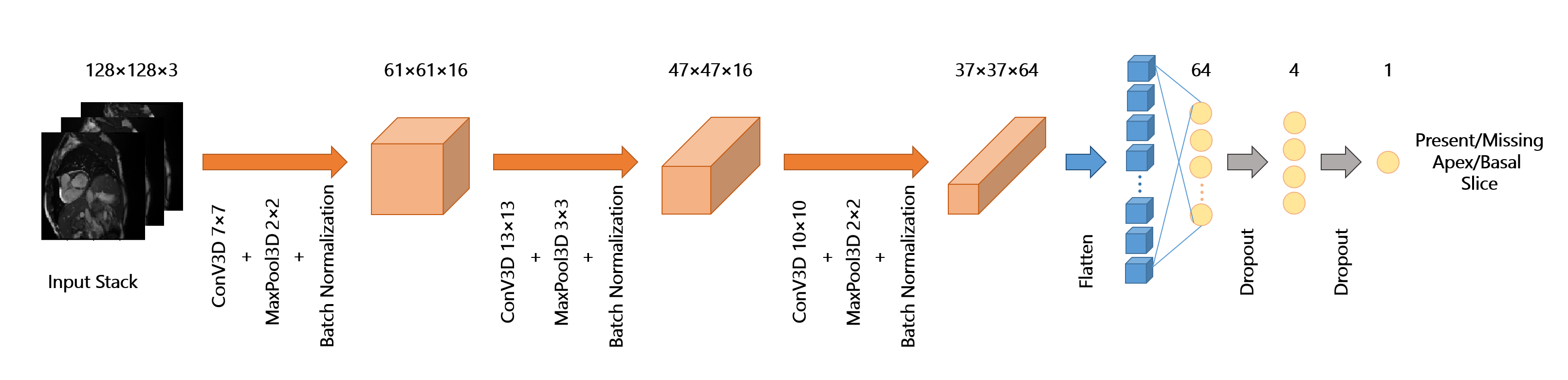}}
\caption{The baseline model's architecture and parameter settings to identify the presence/absence of basal/apical slices.}
\label{fig2}
\end{figure*}

Since to examine the complete cardiac coverage, it is necessary to analyze the volumetric data, a deep 3D convolutional network is used at this stage. Using 3D convolutional kernels can extract each image's desired spatial features and obtain the appropriate related features between successive slices. Therefore, 3D features in the form of cubes of the same size with a convolutional kernel are extracted from each image stack. By sliding the 3D kernel on these stacks, the feature maps of each convolutional layer are obtained. The network is trained twice, once to identify the presence/absence of basal slices and once to identify the presence/absence of apical slices. At the end of the training process, the trained models for apex and basal slice detection are stored as \(\omega_{apex}\) and \(\omega_{basal}\). The architectural details of the baseline model are shown in Fig. \ref{fig2}. In this figure, the inputs are passed through a 3D convolutional layer, a pooling layer for down-sampling, and a batch normalization layer. This process is repeated three times to extract the feature from the input stacks. The extracted feature vector is then passed through the three fully connected layers for classification.

\subsubsection{Explainable Deep Visual Salient Region Detection Model}
\begin{figure*}[!t]
\centerline{\includegraphics[width=0.77\textwidth]{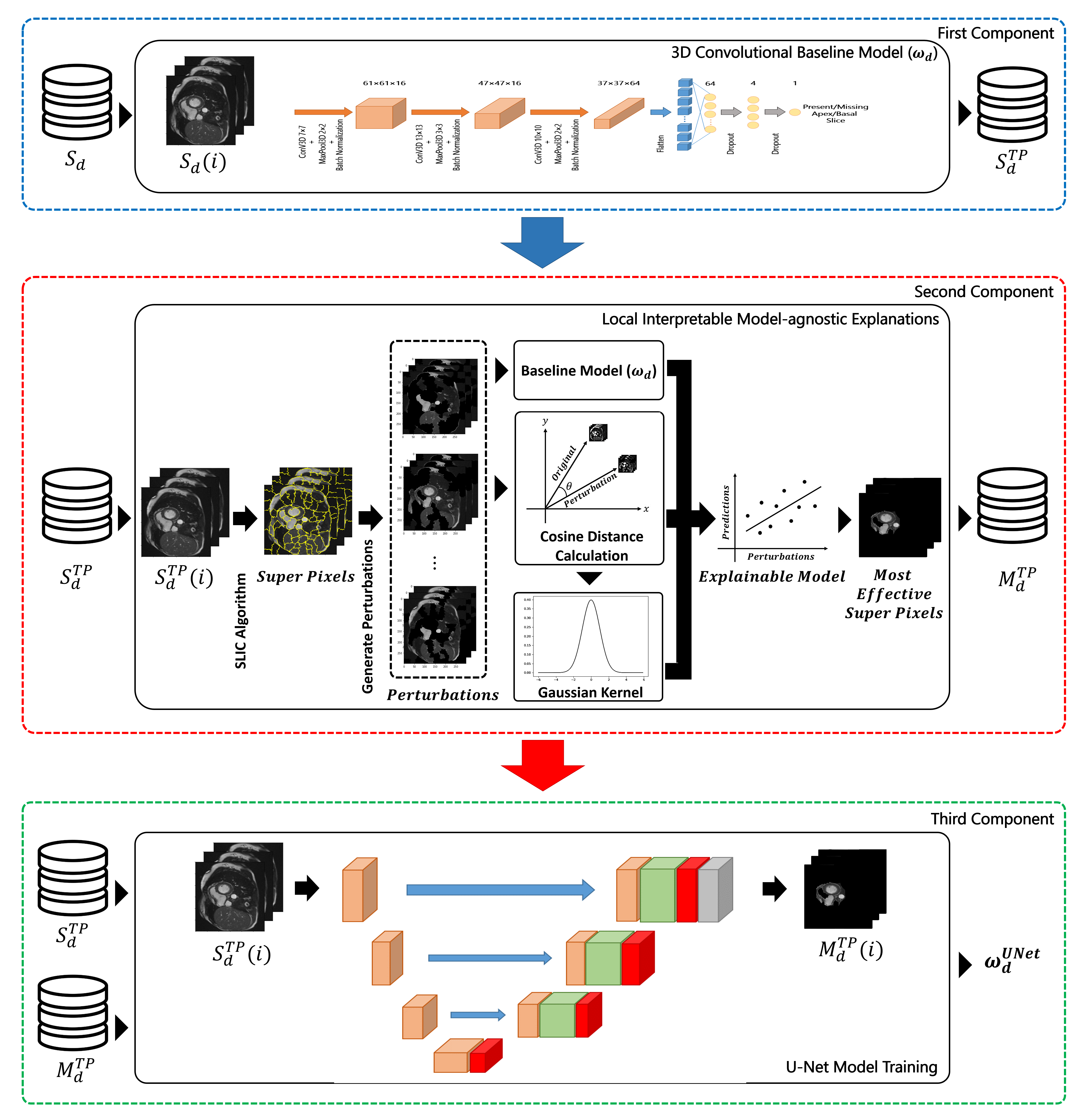}}
\caption{An overview of the proposed explainable deep visual salient region detection model. The first component is related to the baseline model for detecting the presence/absence of basal/apical slices. The second component interprets the baseline model to extract the most significant super-pixel in the input stacks to produce corresponding masks. The third component is used for extracting salient regions.}
\label{fig3}
\end{figure*}

In this section, we propose an innovative model for salient region detection of input stacks. An overview of this proposed model is shown in Fig. \ref{fig3}. The proposed model is supposed to receive the input stacks and finally extract the salient regions. This model consists of the following three components:
\begin{enumerate}
    \item Component for detecting the presence/absence of basal/apical slices
    \item Component of interpreting the baseline model to extract the most significant super-pixel in the input stacks to produce corresponding masks
    \item Component for extracting salient regions
\end{enumerate}

Here, the idea is to use the first component to collect a set of full-coverage stacks (positively labelled stacks). Then, using the second component to detect the most significant super-pixels in the positively labelled stacks as the corresponding mask. Finally, we use the third component to extract the salient region of coming stacks.

The first component is the trained baseline model to identify the presence/absence of basal/apical slices. \(S_{apex}\) and \(S_{basal}\) are tested using baseline models stored in the form of \(\omega_{apex}\) and \(\omega_{basal}\), and all positive stacks that these models correctly classified are stored as \(S^{TP}_{apex}\) and \(S^{TP}_{basal}\) datasets. We use \(S^{TP}_{apex}\) and \(S^{TP}_{basal}\) to investigate the baseline model's interpretability and produce the masks required for the third component. Therefore, \(S^{TP}_{apex}\) and \(S^{TP}_{basal}\) datasets are formed by testing the baseline model to be used as the input of the second component.

Ensuring the performance of black-box machine learning models to understand their predictions is one of the issues that has been considered in recent years \cite{gilpin2018explaining}. Interpretation of machine learning models can reveal features that have had a more significant impact on model prediction, and the final prediction depends heavily on these features. In medical applications, the need to interpret the models is much more important than other areas, given that their predictions are to be used in the treatment of patients \cite{vellido2020importance}. The second component examines the interpretability of the baseline model by receiving \(S^{TP}_{apex}\) and \(S^{TP}_{basal}\) datasets and determines the most effective regions of input stacks that have had the most significant impact on the correct classification of the baseline model. The Local Interpretable Model-agnostic Explanations (LIME) technique \cite{ribeiro2016should} is used to check the model interpretability and produce the corresponding masks of the input stacks. These masks represent the salient regions of the input stacks identified during the model interpretability checking process. Thus, \(S^{TP}_{apex}\) and \(S^{TP}_{basal}\) datasets are the second component's input, and the \(M^{TP}_{apex}\) and \(M^{TP}_{basal}\) sets are generated as the corresponding masks at the end of this step.

To generate explanations for \(S^{TP}_{apex}\) and \(S^{TP}_{basal}\) datasets, the images in each stack must be segmented into a set of super-pixels using a segmentation algorithm. The segmentation of images into super-pixels is done by leveraging the simple linear iterative clustering (SLIC) algorithm \cite{achanta2012slic} with a low computational cost while having a suitable segmentation quality. By turning on and off some super-pixels, various perturbations of the input image are generated. Each perturbation is a Boolean vector of \(1\) for turned on super-pixels and \(0\) for turned off super-pixels. The generation of perturbations is based on a binomial distribution. The baseline CNN model (\(\omega_{apex}\) or \(\omega_{basal}\)) is used to identify the class associated with each perturbed stacks. The baseline CNN model prediction results for perturbed stacks are obtained in the form of numbers between zero and one, which indicates the probability of belonging to the positive class. Perturbation can now be used as input features and predictions for the positive class as desired output to fit a linear interpretable model. Before fitting the linear model, a weight for each perturbation is also calculated based on the cosine distance between the original image as a perturbation with all turned on super-pixels and the generated perturbation. Equation \eqref{eq6} shows how to calculate the cosine distance.

\begin{equation}
Cosine\_Distance=1-Cosine\_Similarity
\label{eq6}
\end{equation}

\noindent where \(Cosine\_Similarity\) can be determined using \eqref{eq7}.

\begin{equation}
\begin{split}
Cosine\_Similarity(Original\_Image, Perturbation)= \\
=\cos(\theta)=\frac{Original\_Image \cdot Perturbation}{\lVert  Original\_Image\rVert \lVert Perturbation\rVert}
\end{split}
\label{eq7}
\end{equation}

After calculating the cosine distance, a Gaussian kernel function converts the distance to a weight with a number between zero and one. Equation \eqref{eq8} defines this Gaussian kernel function.

\begin{equation}
Weight=\sqrt {\exp({-\frac{{Cosine\_Distance}^2}{{Kernel\_width}^2}})}
\label{eq8}
\end{equation}

A weighted regression model is then fitted using the generated information, including perturbations and corresponding predictions' values and weights. We now consider the coefficients of this weighted regression model for each super-pixel. These coefficients indicate how strong the super-pixels were in predicting the input stack as a positive sample. By sorting these coefficients for all super-pixels, it is determined which was the most effective super-pixel for predicting the positive class. The super-pixel with the highest coefficient is the area that has had the most impact on the prediction of the baseline CNN model and is returned as the model explanation by LIME. By examining the areas obtained by LIME for the input stacks, it can be seen whether the model has focused on the convenient features for prediction. If the model works based on the convenient features, it can be said that the model is reliable. The most effective super-pixels extracted as a mask corresponding to each stack are stored as \(M^{TP}_{apex}\) and \(M^{TP}_{basal}\) sets.

The proposed model's third component is related to training an attention-based U-Net model to extract salient regions of input stacks. By considering the most effective super-pixels in prediction as a ground-truth (\(M^{TP}_{apex}\) and \(M^{TP}_{basal}\)) and the corresponding positive input stacks (\(S^{TP}_{apex}\) and \(S^{TP}_{basal}\)) to detect the presence/absence of apical/ basal slices, a pair can be achieved for 3D attention U-Net network \cite{islam2019brain} training. Repeating this process for all positive stacks forms a set of pairs that, by training a U-Net network based on this dataset, can lead to a model that extracts the salient regions of the input stacks, which are important for our baseline CNN model. This 3D U-Net model uses a set of encoders for feature extraction. The extracted feature maps by encoders are given in the form of skip connections as the input of decoders that are integrated with 3D attention modules. Combining decoders with 3D attention modules can improve segmentation results. More details about the architecture of this U-Net model can be found in \cite{islam2019brain}. Finally, after completing the training and evaluation of the 3D attention U-Net, this network will be able to determine the salient regions of each stack. The training and evaluation of this network is done in two separate stages for salient region detection in the \(S^{TP}_{apex}\) and \(S^{TP}_{basal}\) datasets, which leads to two sets of weights, \(\omega^{UNet}_{apex}\) and \(\omega^{UNet}_{basal}\). Algorithm \ref{alg1} provides the pseudo-code related to salient region detection based on the proposed model.

\begin{algorithm*}
\caption{Salient Region Detection Model}
\label{alg1}
    \begin{algorithmic}[1]

      \REQUIRE $S_d\ \mathrm{ is\ input\ Dataset}, \omega_d\ \mathrm{is\ Baseline\ Model's\ weights\ where} \ d \in \{apex,basal\}$
      \ENSURE $\omega_d^{UNet}\ \mathrm{is\ 3D\ attention\ UNet\ weights}$
      \newline
      \STATE $S_d^{TP}=\{\} \hspace*{19em} \mathrm{\triangleright\ First\ Component:} $
      \FOR{$S_d(i)\in S_d$}
      \STATE  $Prediction\leftarrow \mathrm{Predict\ Class\ using\ Baseline\ Model\ given}\ S_d(i)\ \mathrm{and}\ \omega_d $
        \IF {$Prediction\ \mathrm{is\ true\ positive}$}
           \STATE $S_d^{TP}\leftarrow S_d^{TP}\ \cup\ S_d(i)\ $
        \ENDIF
      \ENDFOR \newline
      \STATE $M_d^{TP}=\{\} \hspace*{18em} \mathrm{\triangleright\ Second\ Component:}$
      \FOR{$S_d^{TP}(i)\in S_d^{TP}$}
      \STATE $Super\_Pixels \leftarrow \mathrm{Segment}\ S_d^{TP}(i)\ \mathrm{using\ SLIC\ Alg.} $
      \STATE $Perturbations \leftarrow \mathrm{Generate\ Perturbations\ for}\ Super\_Pixels $
      \STATE $Predictions \leftarrow \mathrm{Predict\ Class\ using\ Baseline\ Model\ given}\ Perturbations\  \mathrm{and}\ \omega_d $
      \STATE $Distances \leftarrow \mathrm{Calculate\ Cosine\ distance\ between}\ S_d^{TP}(i)\ \mathrm{and}\  Perturbations\ \hspace*{8em} \mathrm{\triangleright\ Eq.\ (6),(7)}$
      \STATE $Weights \leftarrow \mathrm{Apply\ Gaussian\ kernal\ on}\ Distances\ \hspace*{19.6em} \mathrm{\triangleright\ Eq.\ (8)}$
      \STATE $Explainable\_Model \leftarrow \mathrm{Fit\ a\ weighted\ linear\ regression\ using}\ Perturbations,\ Predictions\ \mathrm{and}\ Weights$
      \STATE $Coefficients \leftarrow \mathrm{Coefficients\ of}\ Explainable\_Model $
      \STATE $Top\_feature \leftarrow \mathrm{Maximum\ of}\ Coefficients$
      \STATE $Mask \leftarrow \mathrm{Create\ mask\ for}\ Top\_feature$
      \STATE $M_d^{TP}\leftarrow M_d^{TP}\ \cup\  Mask $
      \ENDFOR \newline
   $\hspace*{23em} \mathrm{\triangleright\ Third\ Component:}$

      \STATE $\omega_d^{UNet}\leftarrow \mathrm{Train\ UNet\ Model\ using}\ M_d^{TP}$
      \newline
      \RETURN $\omega_d^{UNet} \hspace*{17em} \mathrm{\triangleright\ 3D\ attention\ UNet\ weights }$
    \end{algorithmic}
\end{algorithm*}

\subsubsection{Baseline Model Improvement using the Extracted Salient Regions}
After evaluating the salient region detection model and obtaining \(\omega^{UNet}_{apex}\) and \(\omega^{UNet}_{basal}\), we use \(\omega^{UNet}_{apex}\) and \(\omega^{UNet}_{basal}\) to improve the quality of baseline model results. For this purpose, based on what is stated in Algorithm \ref{alg2}, we first predict the class of each sample of the testing set. If the prediction result for the \(i^{th}\) stack shows that the class is positive, we accept the result. Otherwise, when class prediction is negative, we first pass the \(i^{th}\) stack to the salient region detection model (\(\omega^{UNet}_{apex}\) and \(\omega^{UNet}_{basal}\)) and then repeat the prediction using the baseline model for the obtained salient regions. We seek to guide the baseline model in identifying positive stacks using the salient region detection model cascaded with the baseline model. At the end of this process, the baseline model evaluation is performed again to determine the improvement.

\begin{algorithm*}
    \caption{Baseline Model Improvement}
    \label{alg2}
    \begin{algorithmic}[1]

      \REQUIRE $S_d\ \mathrm{ is\ input\ Dataset}, \omega_d\ \mathrm{is\ Baseline\ Model's\ weights},\  \omega_d^{UNet}\ \mathrm{is\ UNet\ Model's\ weights\ where}\ d \in \{apex,basal\}$

      \ENSURE $ \mathrm{Predictions\ of\ Baseline\ Model\ after\ Improvement\ using\ Alg.\ 1} $ \newline

      \FOR{$S_d(i)\in S_d$}
      \STATE  $Prediction(i) \leftarrow \mathrm{Predict\ Class\ using\ Baseline\ Model\ given}\ S_d(i)\ \mathrm{and}\ \omega_d $
        \IF {$Prediction(i)\ \mathrm{is\ Negative}$}
           \STATE $Salient\_Region \leftarrow \mathrm{Extract\ Salient\ Region\ using\ Alg.\ 1\ given}\ S_d(i)\  \mathrm{and}\ \omega_d^{UNet} $
           \STATE $Prediction(i) \leftarrow \mathrm{Predict\ Class\ using\ Baseline\ Model\ given}\ Salient\_Region\ \mathrm{and}\ \omega_d $
        \ENDIF
      \ENDFOR \newline
      \RETURN $Prediction \hspace*{8.5em} \mathrm{\triangleright\ Predictions\ for\ All\ Input\ Stacks} $

    \end{algorithmic}
\end{algorithm*}

\section{Experimental Setup}
\subsection{Dataset Description}
The CMR image database from the UK Biobank is used to perform the intended experiments. CMR image acquisitions of the UK biobank has been performed using a wide clinical bore \(1.5\ T\) MR system (MAGNETOM Aera, Syngo Platform VD13A, Siemens Healthcare, Erlangen, Germany) with an \(18\) channel anterior body surface coil (\(45\ mT/m\) and \(200\ T/m/s\) gradient system). All acquisitions have been made using balanced steady-state free precession (bSSFP) MRI sequences to achieve complete heart coverage. The spatial resolution of short-axis cine CMR images is \(1.8\times1.8\ mm\) with a slice thickness of \(8\ mm\) and a slice gap of \(2\ mm\). Each volumetric sequence contains approximately \(50\) cardiac phases, and each imaging volume includes about ten slices to cover the heart completely. For more information on UK Biobank CMR imaging protocols, refer to \cite{petersen2015uk}.

\subsection{Model Training and Hyper-Parameter Setting}
A computer system equipped with an NVIDIA GeForce GTX TITAN X GPU and 16GB of RAM was used to train and evaluate the models proposed in this study. The stochastic gradient descent \cite{ruder2016overview} and binary cross-entropy \cite{wang2020comprehensive} are used as optimizer and loss function to learn the baseline model. The learning rate was considered to be \(0.001\). Learning the baseline model is done in 50 epochs with batch\_size\(=8\) and input stack size \(128\times128\times3\). The performance evaluation is conducted using 5-fold cross-validation \cite{wong2019reliable}. Besides, the hyper-parameters are obtained based on ablation studies.

The SLIC algorithm is leveraged to find super-pixels to investigate the baseline model interpretability. For this purpose, the number of segments, compactness and maximum iteration were considered \(25\), \(0.3\) and \(1000\), respectively. The assumed kernel width in \eqref{eq8} is \(0.25\). These values have been selected after several investigations and consultations with physicians. All the stacks belonging to the positive class of datasets correctly predicted by the baseline model are used to learn the 3D attention U-Net. Details of this network implementation are given in \cite{islam2019brain}. The codes related to the models developed in this study will be released on GitHub after the article is published.

Data augmentation was used to enlarge the training set to improve the baseline model's performance and prevent overfitting \cite{shorten2019survey}. A set of geometric transformations and brightness modifications is used to augment the data. Due to the use of specific image acquisition protocols of the UK biobank, there are few significant variabilities in the images. Therefore, the model must be reliable concerning images' possible geometric and brightness changes. A set of rotational transformations at angles between \(-45^{\circ}\) to \(+45^{\circ}\), horizontal and vertical flips, and brightness changes with random values between \(0\) and \(1\) were considered to augment the data. This data augmentation process doubles the number of training data.

\subsection{Experiment Designs and Performance Metrics}
Several experiments are performed to evaluate the performance of the proposed model. First, two experiments are performed to assess the 3D convolutional baseline model to detect the presence/absence of basal/apical slices. These experiments provide a benchmark for comparing improved results using the proposed salient region detection model. Experiments are done in 5-fold cross-validation and by dividing the dataset into 80\% for training and 20\% for testing. Then in other experiments, the performance of the proposed salient region detection model to improve the baseline model is examined.

After testing the baseline model in two basal/apical slices detection experiments, all incorrectly classified positive class input stacks are given to the proposed salient region detection model and then retested. These experiments investigate whether the proposed salient region model can help improve the baseline model's performance. The results of the U-Net model evaluation are also presented in the results section.

Five evaluation metrics, including accuracy (ACC), precision (PR), recall (RE), F-measure, and area under the ROC curve (AUC), are used to evaluate the proposed model. Equations \eqref{eq9} to \eqref{eq13} show how these metrics are calculated. Also, the Dice coefficient and Jaccard index, given in \eqref{eq14} and \eqref{eq15}, are leveraged to assess the U-Net model.

\begin{equation}
ACC=\frac{TP+TN}{TP+FP+TN+FN}
\label{eq9}
\end{equation}

\begin{equation}
PR= \frac{TP}{TP+FP}
\label{eq10}
\end{equation}

\begin{equation}
RE= \frac{TP}{TP+FN}
\label{eq11}
\end{equation}

\begin{equation}
F-measure= \frac{2\times PR\times RE}{PR+RE}
\label{eq12}
\end{equation}

\begin{equation}
AUC= \int_0^1 Pr[TP] (v) \mathrm{d}v
\label{eq13}
\end{equation}

\begin{equation}
Dice\_Coefficient=\frac{2\times TP}{FN+(2\times TP)+FP}
\label{eq14}
\end{equation}

\begin{equation}
Jaccard\_Index=\frac{TP}{TP+FN+FP}
\label{eq15}
\end{equation}

\noindent where \(TP\), \(FP\), \(FN\) and \(TN\) are true positive, false positive, false negative, and true negative. AUC indicates the overall success of an experiment where \(Pr[TP]\) is a function of \(v=Pr[FP]\).

\section{Results}
\subsection{3D Convolutional Baseline Model}
In the first step, to create a benchmark for comparison, stacks related to detecting the presence/absence of basal/apical slices are used in two separate experiments to evaluate the 3D convolutional baseline model. In the first experiment, 10898 stacks, including 5449 positive and 5449 negative stacks to maintain class balance, are used to assess the presence/absence of basal slice. 80\% of this database is leveraged for training and 20\% for testing. Data augmentation is done to achieve better results so that the training data is doubled in size. The experiment results with the same setting to identify the presence/absence of basal/apical slice are given in Table \ref{tab1}.

\begin{table*}[!htp]\centering
\caption{Evaluation results of the 3D convolutional baseline model to detect the presence/absence of apical/basal slices based on accuracy, precision, recall, F-measure and AUC metrics. The results are obtained using 5-fold cross-validation. The results of each fold are mentioned for more clarity.}
\label{tab1}
\begin{tabular*}{\textwidth}{lcccccccccc}\toprule

\multirow{2}{*}{\textbf{}}
&\multicolumn{2}{c}{\textbf{Accuracy (\%)}} &\multicolumn{2}{c}{\textbf{Precision (\%)}} &\multicolumn{2}{c}{\textbf{Recall (\%)}} &\multicolumn{2}{c}{\textbf{F-measure (\%)}} &\multicolumn{2}{c}{\textbf{AUC (\%)}} \\
\cmidrule(lr){2-3} \cmidrule(lr){4-5} \cmidrule(lr){6-7} \cmidrule(lr){8-9} \cmidrule(lr){10-11}
&Apical &Basal &Apical &Basal &Apical &Basal &Apical &Basal &Apical &Basal \\
\cmidrule(lr){2-3} \cmidrule(lr){4-5} \cmidrule(lr){6-7} \cmidrule(lr){8-9} \cmidrule(lr){10-11}
Fold \#1 &96.00 &95.68 &96.70 &97.03 &95.39 &94.14 &96.04 &95.56 &96.01 &95.66 \\
Fold \#2 &94.25 &96.87 &93.81 &96.27 &94.60 &97.51 &94.21 &96.88 &94.25 &96.87 \\
Fold \#3 &94.02 &95.86 &94.11 &95.61 &93.58 &96.14 &93.85 &95.87 &94.01 &95.86 \\
Fold \#4 &94.76 &96.64 &92.24 &95.78 &97.33 &97.48 &94.72 &96.62 &94.84 &96.65 \\
Fold \#5 &93.52 &96.18 &95.97 &95.26 &91.54 &97.40 &93.70 &96.32 &93.63 &96.15 \\
\cmidrule(lr){2-3} \cmidrule(lr){4-5} \cmidrule(lr){6-7} \cmidrule(lr){8-9} \cmidrule(lr){10-11}
Avg±SD & 94.51±0.95 &96.25±0.51 &94.57±1.78 &95.99±0.69 &94.49±2.15 &96.53±1.46 &94.50±0.94 &96.25±0.54 &94.55±0.93 &96.24±0.51 \\
\bottomrule
\end{tabular*}
\end{table*}

\subsection{Salient Region Detection Model}
\begin{figure}[!t]
\centerline{\includegraphics[width=0.7\columnwidth]{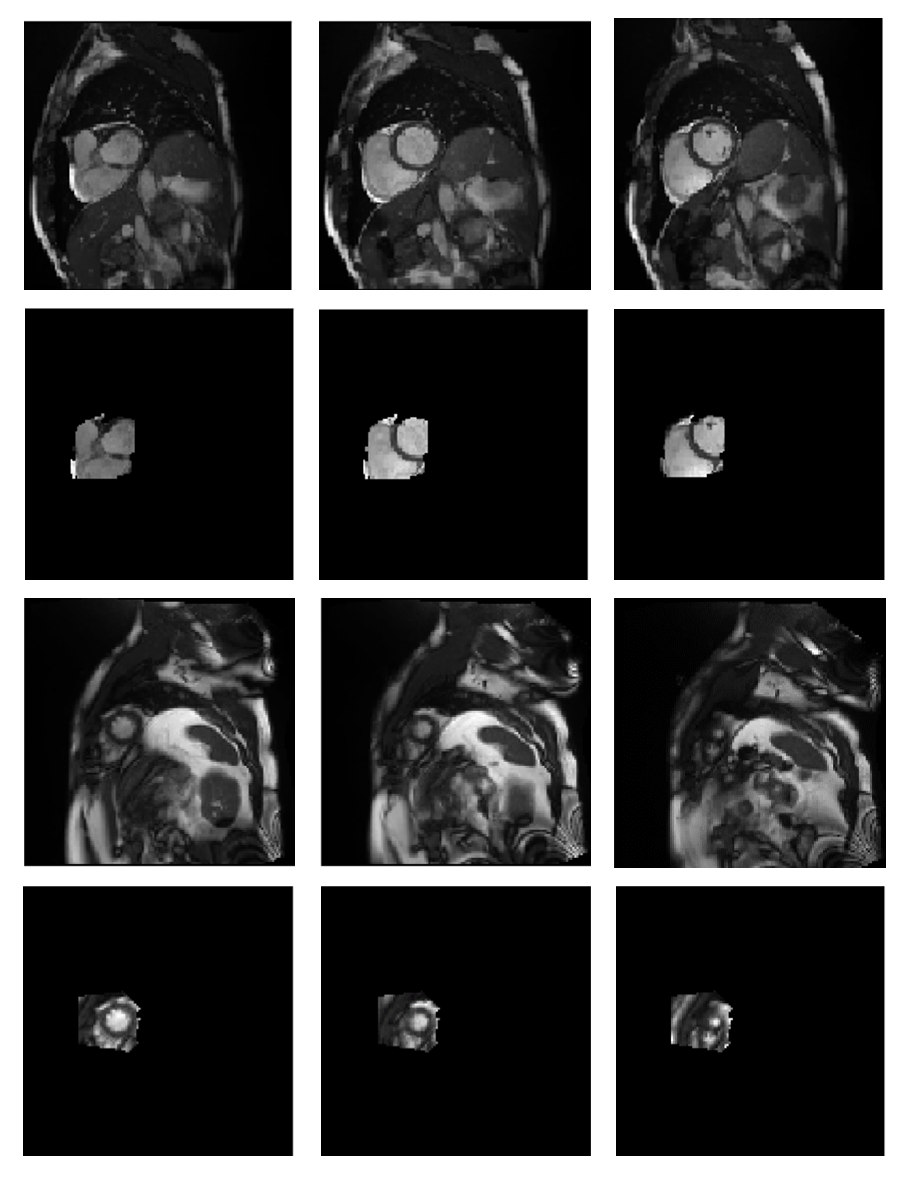}}
\caption{An example of interpretability control results of the baseline model. The first and third rows show the input stacks to the baseline models of the presence detection of basal and apex slices, respectively. The second and fourth rows show the most effective super-pixels extracted by the interpretability control process for the input stacks.}
\label{fig4}
\end{figure}

After training the 3D convolutional baseline model and performing the steps related to examining the interpretability of this model, and training the two U-Net models based on the most effective super-pixel obtained to identify the basal/apical slice, the proposed model can be used to extract the salient region of new stacks. If the most effective super-pixel in the prediction contains an area that includes the heart, it can be said that the baseline model has made a decision based on the correct region of the input stacks. An example of the most effective super-pixels in predicting basal/apical slice presence obtained during the interpretability control process of the baseline model is shown in Fig. \ref{fig4}. As shown in Fig. \ref{fig4}, the super-pixels that have had the most significant impact on classification include the area where the heart is visible. An experienced cardiologist subjectively monitored the baseline model's interpretability control process results to ensure this issue. Accordingly, over 95\% of the results contain areas that include the heart. This indicates that the baseline model is interpretable and works well.

U-Net model training is done using the interpretability control results of the baseline model. Accordingly, two U-Net models are trained to extract salient regions in basal stacks and the other for apex stacks. Table \ref{tab2} shows the training results of these two U-Net models.

\begin{table}[!htp]\centering
\caption{Results of U-Net model training for salient region detection in basal and apical 3D stacks.}
\label{tab2}
\scriptsize
\begin{tabular}{lccc}\toprule
&Dice Score (\%) &Jaccard Index (\%) \\\midrule
Apex U-Net Model &66.10 &50.17 \\
Basal U-Net Model &83.20 &71.02 \\
\bottomrule
\end{tabular}
\end{table}

\subsection{The Baseline Model Improvement}
In the other experiment, the improvement of the 3D convolutional baseline model is investigated using the proposed model of salient region detection. For this purpose, first, the test set is given to the trained baseline models including \(\omega_{apex}\) and \(\omega_{basal}\). For all positive input stacks classified as negative by the baseline models, we extract the salient regions by the proposed model of salient region detection. We retested the baseline model by replacing the original stacks of positive stacks, classified as negative, with the salient regions. In this experiment, positive stacks that are correctly classified remain intact. The dataset size and other settings are similar to the first experiment. The results of the baseline model improvement using the proposed salient region detection model in 5-fold cross-validation are given in Table \ref{tab3}.

\begin{table*}[!htp]\centering
\caption{Evaluation results of the baseline model improvement using the proposed salient region detection model for apical/basal stacks based on accuracy, precision, recall, F-measure and AUC metrics. The results are obtained using 5-fold cross-validation. The results of each fold are mentioned for more clarity.}
\label{tab3}
\begin{tabular}{lccccccccccc}\toprule
\textbf{} &\multicolumn{2}{c}{\textbf{Accuracy (\%)}} &\multicolumn{2}{c}{\textbf{Precision (\%)}} &\multicolumn{2}{c}{\textbf{Recall (\%)}} &\multicolumn{2}{c}{\textbf{F-measure (\%)}} &\multicolumn{2}{c}{\textbf{AUC (\%)}} \\
\cmidrule(lr){2-3} \cmidrule(lr){4-5} \cmidrule(lr){6-7} \cmidrule(lr){8-9} \cmidrule(lr){10-11}
\textbf{} &Apical &Basal &Apical &Basal &Apical &Basal &Apical &Basal &Apical &Basal \\
\cmidrule(lr){2-3} \cmidrule(lr){4-5} \cmidrule(lr){6-7} \cmidrule(lr){8-9} \cmidrule(lr){10-11}
Fold \#1 &97.47 &96.55 &96.78 &97.07 &98.27 &95.90 &97.52 &96.48 &97.46 &96.54 \\
Fold \#2 &95.45 &97.42 &95.64 &96.30 &95.38 &98.61 &95.51 &97.44 &95.45 &97.52 \\
Fold \#3 &95.12 &96.50 &95.44 &95.66 &94.92 &97.42 &95.18 &96.53 &95.13 &96.50 \\
Fold \#4 &95.72 &97.01 &93.84 &95.81 &98.00 &98.22 &95.87 &97.00 &95.69 &97.00 \\
Fold \#5 &94.85 &96.92 &96.44 &95.32 &93.29 &98.83 &94.84 &97.04 &94.87 &96.87 \\
\cmidrule(lr){2-3} \cmidrule(lr){4-5} \cmidrule(lr){6-7} \cmidrule(lr){8-9} \cmidrule(lr){10-11}
Avg±SD &95.72±1.03 &96.88±0.38 &95.63±1.14 &96.03±0.68 &95.97±2.12 &97.80±1.19 &95.78±1.04 &96.90±0.40 &95.72±1.02 &96.89±0.41 \\
\bottomrule
\end{tabular}
\end{table*}

\subsection{Comparison to Other Related Methods}
The baseline model presented in this study and its improvement outperforms the previous studies. The results of these comparisons are shown in Table \ref{tab4}. First, the conventional 3D CNN model presented in the study of Zhang et al. \cite{zhang2018automatic}, and then their improved model, which is based on the use of a layer called Fisher-Discriminative, was evaluated with the \(S_{apex}\) and \(S_{basal}\) datasets. This table also presents the results of our proposed baseline model and the results related to improving the baseline model using the proposed salient region detection model for comparison. Similar settings were used in the experiments to make the results comparable. The results clearly show the superiority of the proposed models over the previous state-of-the-art method.

\begin{table*}[!htp]\centering
\caption{Performance comparison of the proposed baseline model and its improvement with the previously presented models.}
\label{tab4}
\begin{tabular}{lccccccc}\toprule
\multirow{2}{*}{\textbf{}} &\multicolumn{2}{c}{\textbf{Accuracy (\%)}} &\multicolumn{2}{c}{\textbf{Precision (\%)}} &\multicolumn{2}{c}{\textbf{Recall (\%)}} \\
\cmidrule(lr){2-3} \cmidrule(lr){4-5} \cmidrule(lr){6-7}
&Apex &Basal &Apex &Basal &Apex &Basal \\
\cmidrule(lr){2-3} \cmidrule(lr){4-5} \cmidrule(lr){6-7}
3D CNN Baseline Model by Zhang et al. \cite{zhang2018automatic} &91.92 &92.11 &92.49 &92.52 &92.14 &92.52 \\
Fisher-Discriminative CNN Model by Zhang et al. \cite{zhang2018automatic} &92.51 &92.61 &92.90 &92.91 &92.89 &93.08 \\
Proposed Baseline Model &94.51 &96.25 &94.57 &95.99 &94.49 &96.53 \\
Proposed Model Improvement &\textbf{95.72} &\textbf{96.88} &\textbf{95.63} &\textbf{96.03} &\textbf{95.97} &\textbf{97.80} \\
\bottomrule
\end{tabular}
\end{table*}

\section{Discussion}
In this study, we examined one of the significant challenges in CMR imaging. This challenge is the quality control of CMR images from the perspective of complete cardiac coverage, which affects cardiac functional and volumetric measurements. For this purpose, a 3D convolutional baseline model is improved using a proposed explainable salient region detection model. Short-axis cine CMR images can be examined using the method proposed in this study with considerable accuracy in terms of complete coverage of the heart. Besides, a novel model was proposed for the salient region extraction of images that could be used in other studies.

CMR imaging is a non-invasive technique without ionizing radiation, which is acquired to study the prognosis and diagnosis of cardiovascular disease. CMR images can be used to examine a wide range of heart disorders that require LV volume and shape assessment. Some of the most important of these diseases are hypertrophic cardiomyopathy (HCM) \cite{quarta2018cardiovascular}, dilated cardiomyopathy (DCM) \cite{pirruccello2020analysis}, valvular heart disease \cite{mathew2018role}, congenital heart defects (CHDs) \cite{muscogiuri2017utility}, infectious endocarditis (IE) \cite{bruun2014cardiac} and so on. Due to the high incidence of these diseases and the increasing use of CMR imaging, the need for quality control of these images is very much felt. The assessment of complete cardiac coverage in CMR images is one of the basic quality controls in diagnosing diseases mentioned. Besides, this quality control process must be automated in extensive population studies such as the UK Biobank because subjective evaluation will be practically laborious and time-consuming \cite{petersen2013imaging}. Therefore, in this study, we presented an automated model to control the complete heart coverage in over 6,200 participants in the UK Biobank study, which has the highest number of participants compared to its previous studies. In addition to being an automated model, it needs to be functionally reliable. The model proposed in this study outperforms its prior models.

The baseline model's performance shows a significant improvement over previous 3D CNN models due to batch normalisation. Batch normalisation can improve the accuracy and speed up the network training. Covariate shift occurs if the input distribution to the network changes and internal covariate shift also occurs between layers of deep neural networks. Modifying the input values of successive layers by normalising the activations to a mean of zero and a standard deviation of one reduces the dependence of the gradients on the scale of the parameters or initial values and decreases the covariate shift \cite{bjorck2018understanding}. This issue results in faster convergence and better generalisation and can improve accuracy. The addition of batch normalisation capability after 3D CNN layers achieved these goals in the current study.

Another advantage of the proposed model is using salient image region detection to improve the baseline model's performance. Based on the experiments performed, it can be clearly seen that in situations where neural networks can not make the right decision, they can be helped to complete the task correctly by extracting the salient regions of the images. The proposed salient region extraction model can cause the network to focus on the region of interest (ROI) by removing non-informative areas from the input stacks. Given that we want the proposed model to be fully automatic, finding the salient image regions using the proposed model can be done correctly even in geometric variations in the position of the organs. However, suppose the ROI is selected consistently from a specific image part. In that case, the heart may not be in the ROI due to positional variations, which makes it necessary to examine subjectively, which contradicts the automated nature of the model.

One of the challenging issues in classification is reducing the number of misclassified samples. In this study, we proposed a salient region detection model to increase the model's accuracy by reducing the number of misclassified samples to examine the complete coverage of the heart. Machine learning models may not correctly classify some training set instances even after completing the model training. To further analyse, we decided to examine whether the proposed model can also tackle the problem of misclassified samples from the training set after model training. For this purpose, Algorithm \ref{alg2} was performed for all training data. After testing the baseline models to identify the presence/absence of basal/apical slices using  \(S_{basal}\) and \(S_{apex}\), 381 and 473 stacks are misclassified by the basal and apical baseline models, respectively. Using the proposed salient region detection algorithm for positive stacks reduces the number of false samples to 253 and 303 for basal and apical baseline models, respectively. 34\% and 35.94\% of false samples are classified correctly using the proposed method for \(S_{basal}\) and \(S_{apex}\), respectively. The results after applying Algorithm \ref{alg2} to \(S_{basal}\) and \(S_{apex}\) are given in Table \ref{tab5}.

\begin{table}[!htp]\centering
\caption{Results of Algorithm \ref{alg2} application for  \(S_{basal}\) and \(S_{apex}\) datasets using the baseline models of basal/apical slice detection.}
\label{tab5}
\begin{tabular}{lcccc}\toprule
&\textbf{Accuracy (\%)} &\textbf{Precision (\%)} &\textbf{Recall (\%)} \\\midrule
Apical Slice &97.22 &96.00 &98.55 \\
Basal Slice &97.68 &96.93 &98.48 \\
\bottomrule
\end{tabular}
\end{table}

\section{Conclusion}
This study examines the heart's complete coverage in short-axis cine CMR images of large population image studies. The proposed model is a cascade detector consisting of a 3D CNN baseline model and an explainable salient region detection model to reduce false negatives. The proposed method of this study can automatically analyse the input stacks of short-axis cine CMR images and guide the imaging team when the input stacks do not have complete coverage of the heart from the apex to the basal. This method can also control the quality of bulk population datasets such as the UK Biobank. The baseline and improved model results show the superiority of the proposed method over the state-of-the-art studies. In addition to identifying the complete coverage of the heart in CMR images, a salient region detection model was proposed that can be used in future studies to extract the salient regions of the images and reduce the models' false negatives. Upcoming works could address other CMR image quality controls.

\section*{Acknowledgment}
This study has been conducted using the UK Biobank CMR dataset under Application 11350. Special thanks to Dr Azar Ejmalian (Assistant Professor at Iran University of Medical Sciences) for providing constructive feedbacks.

\bibliographystyle{IEEEtran}


\end{document}